\documentclass[a4paper,UKenglish]{lipics}
\usepackage{url,graphics,amsfonts,amssymb,tikz,verbatim}
\usepackage{amsmath}
\usepackage[ruled, vlined]{algorithm2e}
\usepackage{tcolorbox}
\usepackage{bbm}
\usetikzlibrary{calc}
\usetikzlibrary{patterns}
\usepackage{algpseudocode}

\title{Parameterized Algorithms for Red-Blue Weighted Vertex Cover on Trees}
\author{Vishnu Veerathu$^{1}$}
\author{Yogesh Tripathi$^{1}$}
\affil{Department of Computer Science and Engineering,\\Indian Institute of Technology Madras, India.\\
	\texttt{\{cs16b030,cs16b044\}@smail.iitm.ac.in}}

\begin{document}
\maketitle

\begin{abstract}

\textproc{Weighted Vertex Cover} is a variation of an extensively studied NP-complete problem, \textproc{Vertex Cover}, in which we are given a graph, $G = (V,E,w)$, where function $w:V \rightarrow \mathbb{Q}^{+}$ and a parameter $k$. The objective is to determine if there exists a vertex cover, $S$, such that $\sum_{v \in S}w(v) \leq k$. In our work, we first study the hardness of \textproc{Weighted Vertex Cover} and then examine this problem under parameterization by $l$ and $k$, where $l$ is the number of vertices with fractional weights. Then, we study the \textproc{Red-Blue Weighted Vertex Cover} problem on trees in detail. In this problem, we are given a tree, $T=(V,E,w)$, where function $w:V \rightarrow \mathbb{Q}^{+}$, a function $c:V \rightarrow \{R,B\}$ and two parameters $k$ and $k_R$. We have to determine if there exists a vertex cover, $S$, such that $\sum_{v \in S}w(v) \leq k$ and $\sum_{\substack{v \in S\\ c(v) = R}}w(v) \leq k_R$. We tackle this problem by applying different reduction techniques and meaningful parameterizations. We also study some restrictive versions of this problem. 
\end{abstract}
\section{Introduction}
The Vertex Cover problem takes a graph, $G=(V,E)$ as an input and a parameter $k$ and the objective is to determine if there exists a set $S \subseteq V$ such that for every edge, $e=\{u,v\}$ in $G$, one of it's endpoints is contained in $S$ and $|S| \leq k$. This problem is one of the Karp's 21 NP-complete problems \cite{Kar72}. This problem is well-studied and FPT algorithms exist using techniques like Iterative compression, Bounded Search Trees and solving Linear Programming version of the problem. The best known algorithm for Vertex Cover is $\mathcal{O}(1.2738^k + kn)$ by Chen et al. \cite{Chen2006ImprovedPU}. Vertex Cover problem  has many real-world applications, including
many in the field of bioinformatics. It can be used in the construction of phylogenetic trees, in phenotype
identification, and in analysis of microarray data\cite{inproceedings}.
\par\vspace{1em}
In this work, we look at a generalization of vertex cover problem, which is Weighted Vertex Cover. In this problem, we are given a graph and there are some weights associated with each graph. The aim is to determine a minimum weight vertex cover where weight of the vertex cover is the sum of weights of all vertices in the vertex cover. We first look at hardness of this problem. Then, we develop Buss kernelization based technique for this problem. We then formulate the problem as a Linear Program and develop kernelization based on it. We also get a 2-approximation solution of this problem from the solution of the Linear Program. We then develop an exact solution for the problem based on branching and study an interesting optimization based on distribution of weights in the graph.
\par\vspace{1em}
We then look into the problem of finding a vertex cover in a tree of size at most $K$, whose vertices are either colored $R$(red) or $B$(blue) and we are allowed to select at most $K_R$ $R$-colored vertices. We present a branching-based algorithm to solve this problem. Finally, we perform a time-complexity analysis of the presented algorithm.
\subsection{Problem statement}
The following two problems are of utmost interest in our work:\\
\newline
\fbox{
\parbox{13.5cm}{
	\textproc{Weighted Vertex Cover}\\
	\textsf{\bfseries Instance:} A graph, $G = (V,E,w)$, where function $w:V \rightarrow \mathbb{Q}^{+}$ and a parameter $k$.\\
	\textsf{\bfseries Compute:} A set $S \subseteq V$ such that $S$ is a vertex cover of $G$ and $\sum_{v\in S} w(v) \leq k$.\\
	\textproc{Bounded Number of Fractional Weights}\\
	\textsf{\bfseries Variation:} The problem is additionally parameterized with $l$, which is the maximum number of vertices in $V$ which have fractional weights.
}
}
\vspace{.5cm}\\
\fbox{
\parbox{13.5cm}{
	\textproc{Red-Blue Weighted Vertex Cover on Trees}\\
	\textsf{\bfseries Instance:} A tree, $T=(V,E,w,c)$, where function $w:V \rightarrow \mathbb{Q}^{+}$ and function $c:V \rightarrow \{R,B\}$ and two parameters $k$ and $k_R$.\\
	\textsf{\bfseries Compute:}A set $S$, such that $S$ is a vertex cover of $T$, $\sum_{v \in S}w(v) \leq k$ and $\sum_{\substack{v \in S\\ c(v) = R}}w(v) \leq k_R$.
}
}
\vspace{.5cm}

\subsection{Related works}
 Minimum Weighted Vertex Cover problem was introduced by Fellows et al. \cite{introtowvc}. A crown reduction based approach to tackle the minimum weighted vertex cover and Nemhauser-Trotter approaches based on relaxed version of NP-problem is explored by Miroslav Chlebíka and Janka Chlebíková in \cite{Chlebik:2008:CRM:1330778.1331075}. The Red-Blue set cover problem study, mainly using Linear Programming formulations and studying the relaxed version of the problem is done by in \cite{Carr:2000:RSC:338219.338271}. This study gave us directions for tackling Red-Blue Vertex Cover. The Red-Blue dominatong set problem is studied in by Venkatesh Raman and Saket Saurabh in \cite{Raman2008}. Some sections from this work were used to guide the ideas in this work.
\section{Preliminaries}
In this article, an undirected graph is represented as $G=(V,E)$, where $V$ is the set of nodes and $E$ is the set of edges. An edge between two nodes $u$ and $v$ is represented by $uv$ or $\{u,v\}$. When defining a weighted graph, we have used the notation, $G = (V,E,w)$, where $w:V\rightarrow \mathbb{Q^{+}}$ is a function with domain as set of vertices and co-domain as set of positive rational numbers. In a graph $G$, the neighborhood of a vertex $v$ is defined as $N(v)$ which is $\{u|uv \in E\}$. A color function on graphs which have some color on each of the vertex, have another function $c$, augmented with definition of $G$. $min(w)$ refers to the minimum weight in the graph, that is, $\min_{v\in V}w(v)$.
\section{\textproc{Weighted Vertex Cover} Kernelizations}

\subsection{Hardness of \textproc{Weighted Vertex Cover}}
\textbf{Theorem.} There exists no $n^{f(k)}$ algorithm for any computable function $f$ over $k$, for \textproc{Weighted Vertex Cover} where $n$ is the number of vertices in the graph, unless P=NP.\\\\
\emph{Proof.} For the sake of contradiction, assume that there exists an $n^{f(k)}$ algorithm for \textproc{Weighted Vertex Cover}. Now, consider a problem instance of \textproc{Vertex Cover}, given a graph $G=(V,E)$ where $|V| = n$, determine if there exists a vertex cover of size at most $k$. Consider the following reduction:\\
Construct a function, $w:V \rightarrow \mathbb{Q}^{+}$ such that $\forall \  u \in V,\ w(u) = \frac{1}{k}$. $(G,1)$ is now a problem instance for \textproc{Weighted Vertex Cover} where we need to find a vertex cover $S$, such that $\sum_{u \in S} w(u) \leq 1$.\\

If there exists an $n^{f(k)}$ algorithm for finding the weighted vertex cover, then this instance can be solved in $n^{f(1)}$. The solution obtained for this instance is also a solution for \textproc{Vertex Cover} instance. Since $f(1) = c$, for some constant $c$, this gives a polynomial time solution for \textproc{Vertex Cover}. This is not possible unless P=NP. Therefore, there is no $n^{f(k)}$ algorithm for \textproc{Weighted Vertex Cover} unless P=NP.

\subsection{Buss Kernelization for \textproc{Weighted Vertex Cover}}
Given an instance of \textproc{Weighted Vertex Cover}, $(G,k)$, apply the following reductions.
\textbf{Reduction 1.}
If $G$ contains an isolated vertex $v$, remove $v$ from $G$. The new instance will be $(G-v,k)$.\\
\textbf{Reduction 2.}
If there exists a vertex $v$ such that $\sum_{u \in N(v)}w(u) > k$, then delete $v$ from the graph and decrement $k$ by $w(v)$. The new instance will be $(G-v,k-w(v))$.\\
Apply the above reductions until neither one can be applied. Since reduction 2 does not hold, $\sum_{u \in N(v)}w(u) \leq k \ \forall v$. Therefore the maximum degree for any vertex is $\frac{k}{min(w)}$. For a $YES$  instance of the problem, there exists a vertex cover $S$ such that $\sum_{v\in S}w(v) \leq k$. It follows that $|S| \leq \frac{k}{min(w)}$. Therefore in a $YES$ instance of the problem, the maximum number of edges that can be present is bounded by $(\frac{k}{min(w)})^2$.\\\\
\textbf{Output of kernelization.}
The problem instance can be concluded to be a $NO$ instance if there are more than $(\frac{k}{min(w)})^2$ edges present in the graph after the above reductions are exhaustively applied. Otherwise, a graph with a maximum of $(\frac{k}{min(w)})^2$ is obtained.

\subsection{Linear Programming Based Formulation}
\subsubsection{Linear Programming Based Kernelization}
This technique is similar to the one in Parameterized Algorithms(2.5)\cite{Cygan:2015:PA:2815661}. A \textproc{Weighted Vertex Cover} instance can be formulated as follows
\begin{align*}
    \text{minimize } \sum_{v\in V} w(v)x_v\\
    \text{with constraints:  } x_u + x_v \geq 1\text{ } \forall uv\in E\\
    0 \leq x_v \leq 1 \text{ }\forall v \in V\\
    x_v\in \textproc{Z} \text{ }\forall v \in V\\
\end{align*}

The last condition, that all $x_v$ must be integers is removed as a relaxation to the problem. The relaxed problem can be solved in polynomial time using Linear Programming techniques. Partition $V$ into three sets as follows.
\begin{align*}
    V_0 = \{v\in V : x_v < 0.5\}\\
    V_1 = \{v\in V : x_v > 0.5\}\\
    V_{\frac{1}{2}} = \{v \in V : x_v = 0.5\}\\
\end{align*}
We know that Nemhauser-Trotter Theorem can be used in case of \textproc{Vertex Cover}\\
\textbf{Nemhauser-Trotter Theorem.} There is a minimum vertex cover $S$ of $G$ such that
\begin{equation*}
    V_1 \subseteq S \subseteq V_1 \cup V_{\frac{1}{2}}
\end{equation*}
\textbf{Proof of Nemhauser-Trotter Theorem(NTT) for Weighted Vertex Cover.} Suppose $S^*$ is a minimum vertex cover of $G$. Let $S$ be defined as below
\begin{equation}
    S = (S^* \setminus V_0) \cup V_1
\end{equation}
We now attempt to show that $S$ is also a minimum vertex cover by contradiction. If $S$ is not a minimum vertex cover, this implies that 
\begin{align*}
    \sum_{v\in A} w(v) < \sum_{v\in B} w(v)\\
\end{align*}
where A,B are $S^*\setminus(V_1 \cup V_{\frac{1}{2}}),S\setminus((V_1 \cap S^*) \cup V_{\frac{1}{2}})$ respectively.\\

Consider the following modification to $x_v$ values obtained from LP methods. For all $v\in B$, decrease the corresponding $x_v$ value by $\epsilon$, and for all $v \in A$, increase the corresponding $x_v$ value by $\epsilon$, for an $\epsilon \leq min_{v \in A\cup B}(|0.5 - x_v|)$. We can now show that the constraints of the relaxed LP problem still hold. From the way $\epsilon$ is defined, we can see that the constraint that $0\leq x_v \leq 1$ $\forall v \in V$ is satisfied. Also note than from the way $\epsilon$ is defined, the splitting of vertices into $V_0,V_1,$ and $V_{\frac{1}{2}}$ remains the same. Increasing any $x_v$ will not cause the constraint that $x_u + x_v \geq 1\text{ } \forall uv\in E$ to be violated. Therefore, it is shown that the constraint holds for all edges for which neither of the vertices it is incident on, belong to $B$.\\

We now consider the different cases in which one of the vertices that the edge is incident on,$u$ or $v$, belong to $B$. If $u,v \in B$, the constraint will be satisfied since $x_u,x_v > 0.5$ even after modifying the values. A similar argument holds for the case in which one of $u$ and $v$ belongs to $B$ and the other belongs to $V_\frac{1}{2}$. This leaves the case in which one vertex belongs to $B$($u$) and the other to $V_0$($v$). $v$ cannot belong to $V_0 \setminus A$ since $S^*$ is a vertex cover and $u$ does not belong to $S^*$, implying that $v \in S^*$. This implies that the constraint $x_u + x_v \geq 1$ holds after modifying the values in this case as well, since the value of $x_v$ was decreased by $\epsilon$ and $x_u$ was increased by $\epsilon$. Thus, we have shown that all constraints are satisfied even after modifying the the $x_v$ for some vertices as specified above. Consider the effect of this modification on the $\sum_{v \in V} w(v)$. $\sum_{v \in V} w(v)$ will change by $(\sum_{v\in A} w(v) - \sum_{v\in B} w(v)) \epsilon \leq 0$, which is a contradiction since the LP method should return the set of values which satisfy the constraints and minimize $\sum_{v \in V} w(v)$.\\
\textbf{Kernelization Using NTT.} There exists a vertex cover which includes all the vertices of $V_1$ and none of the vertices of $V_0$. Kernelize by removing vertices in $V_1$ and $V_0$ and decrease the parameter $k$ by $\sum_{v \in V_1}w(v)$.\\
\textbf{Size Bound of Kernelization}
\begin{align*}
    |V(G^`)| = |V_\frac{1}{2}| = 2\sum_{v\in V_{\frac{1}{2}}}x_v \leq 2\sum_{v \in V}x_v \leq \frac{2k}{min(w)}
\end{align*}
We can also define $d$ as the maximum value such that $\exists w_1,w_2,...w_d$ such that their sum is $\leq k$.
We can then bound the size of the kernel by 2$d$.

\subsubsection{Simple 2-Approximation from NTT solution}
\emph{Theorem.} Let $S = V_{\frac{1}{2}} \cup V_1$. Then, $S$ is a 2-Approximation for \textproc{Weighted Vertex Cover}.

\emph{Proof.} $S$ is obviously a vertex cover because for every edge $e = \{u,v\}\in E$, since $x_u \geq \frac{1}{2}$ and  $x_v \geq \frac{1}{2}$. Therefore, $x_u + x_v \geq 1$ which implies that $e$ is covered.\\

Now, if $S_{OPT}$ is the optimum vertex cover, then $\sum_{u\in S_{OPT}}w(u) \geq \sum_{u\in V} w(u) x_u$ as LP is a relaxed version of the original integer LP problem. Now:
\begin{center}
    \begin{align*}
        \sum_{u\in V}w(u) x_u & \geq \sum_{u\in S} w(u) x_u \\
        &  \geq \frac{1}{2} \sum_{u\in S}w(u) 
\end{align*}
\end{center}

Therefore, $\sum_{u\in S_{OPT}}w(u) \geq \frac{1}{2} \sum_{u\in S}w(u)$. Or equivalently, $2 \sum_{u\in S}w(u) \leq \sum_{u\in S_{OPT}}w(u)$ which proves that $S$ is a 2-approximation.

\section{Weighted Vertex Cover Solutions}

\subsection{Branching Approach for Weighted Vertex Cover Variant}

\subsubsection{Branching Method for Vertex Cover}
This technique is similar to the one in Parameterized Algorithms(3.1)\cite{Cygan:2015:PA:2815661}. It is based on the simple observation that for any vertex $v$, either $v$ or $N(v)$ must be present in a vertex cover. Based on this observation, we "branch on $v$" by creating 2 sub-problems, one in which $v$ has been selected in the vertex cover and is removed and another in which $N(v)$ has been selected in the vertex cover and has been removed. From the earlier observation, it is clear that for this to be a $YES$ instance of the problem, at least one of the sub-problems must be a $YES$ instance. Since, in the worst case, both sub-problems need to be solved to determine whether the current instance of the problem is a $YES$ or a $NO$, the time taken to solve the original problem is the sum of the times taken to solve the 2 sub-problems.\\
\subsubsection{Recursive Formulation}
\textbf{Proposition} The problem can be solved in polynomial time when the maximum degree of the graph is at most 2.\\
Since instances in which the maximum degree is at most 2 can be solved in polynomial time, we consider the instances in which the maximum degree is greater than 2. By branching on the vertex with largest degree(which is at least 3), we obtain the following recursive relations for the time bound function.\\
\begin{equation}
    T(k,l) = a + b
\end{equation}
where
\begin{equation}
    a \in \{T(k-1,l),T(k,l-1)\}
\end{equation}
and 
\begin{equation}
    b \in \{T(k-3,l),T(k-2,l-1),T(k-1,l-2),T(k,l-3)\}
\end{equation}
Each of the 8 equations represent a different case when branching. Here $a$ represents the 2 cases of branching on a fractional weight vertex, and branching on a non-fractional weight vertex. The $b$ term represents the 4 possible cases for the neighbors of the vertex chosen to branch on(Number of fractional weight and non-fractional weight neighbors). If a fractional weight vertex is selected to be part of the vertex cover(and hence removed) $l$ is decremented. If a non-fractional weight vertex is selected as a part of the vertex cover, then $k$ is decremented by 1 since the weight of a non-fractional vertex is at least 1.
\subsubsection{Solving the Recurrence Relation}
Out of the 8 equations above, consider the following 2 equations,
\begin{equation}
    T(k,l) = T(k-1,l) + T(k-3,l)\\
    T(k,l) = T(k,l-1) + T(k,l-3)
\end{equation}

Assuming that the solution is of the form $\alpha^k \beta^l$. By substituting the assumed solution form in the above 2 equations, we can see that $\alpha \geq 1.4656$(by dividing the first equation by $\beta^l$) and that $\beta \geq 1.4656$(by dividing the second equation by $\alpha^k$). Setting $\alpha$ and $\beta$ to be 1.4656 satisfies the other 6 equations as well. Therefore $T(k,l)$, can be estimated to
\begin{align}
    \centering
    T(k,l) = 1.4656^{k+l}
\end{align}
\subsubsection{Optimization Based on Weights Distribution in Special Case}
Consider a special case of the problem. In this variant, $l$ represents the number of fractional weights which are less than 1. The above time bound still holds. This can be shown by slight modifications to the above proof. The vertices which have a fractional weight greater that or equal to 1 can be taken care of by decrementing $k$ instead of decrementing instead of $l$. The same set of recursive relations are obtained, therefore the same solution is obtained. \\
In any weighted vertex cover problem, all the vertex weights and $k$ can be scaled by a constant factor. Therefore the weights and $k$ can be scaled by a constant for a faster running time. However the weights down by a large factor will increase the number of vertices with weight less than 1, therefore increasing $l$.\\
\textbf{Lemma} Scaling all weights and $k$ by $\frac{1}{w_m}$ where $w_m$ is the minimum weight which satisfies $w_m \geq 1$ will not lead to a worse time bound.\\
\emph{Proof.} From the definition of $w_m$, none of the vertices which have a weight greater than or equal to 1 initially will have a weight less than 1, therefore the value of $l$ will not change. $k$ will not increase, since it is being scaled by a value less than or equal to one.Therefore we can consider scaling only by a factor $\frac{1}{w_i}$. Suppose $w_1,w_2,...w_n$ are the weights in ascending order. Then the best obtainable time bound by scaling the weights is
\begin{equation*}
    \text{min } \frac{k}{w_i} + i - 1
\end{equation*}

\section{Red Blue Weighted Vertex Cover on Trees}
\subsection{Red-Blue Vertex Cover on Trees}
Consider the following problem:
\vspace{.5cm}\\
\fbox{
\parbox{13.5cm}{
	\textproc{Red-Blue Vertex Cover on Trees}\\
	\textsf{\bfseries Instance:} A tree, $T=(V,E,c)$, where function $c:V \rightarrow \{R,B\}$ and two parameters $K$ and $K_R$.\\
	\textsf{\bfseries Compute:} A set $S$, such that $S$ is a vertex cover of $T$, $|S| \leq K$ and $\sum_{\substack{v \in V\\ c(v) = R}}1 \leq K_R$.
}
}
\par\vspace{0.7cm}\noindent
\textbf{Theorem.} \textproc{Red-Blue Vertex Cover on Trees} can be solved in $\mathcal{O}(|V|(K_R+K^2)^{K_R})$.\\\\
\emph{Proof.} This proof is divided into two sections. Section 5.1.1 will present a branching based algorithm for the stated problem and Section 5.1.2 will give a bound on the time complexity of the algorithm.
\subsubsection{Branching-based algorithm}
First, we will perform the following reductions on the problem instance.\\
\textbf{Reduction 1.} If there is a vertex $v$, with greater than $K$ neighbors, then add it into the vertex cover, and return the instance $(T,K_R,K-1)$ if $c(v) = B$, return the instance $(T,K_R-1,K-1)$ if $c(v) = R$. If $K-1 <0$ or $c(v) = R$ and $K_R-1 < 0$, return No-instance.
\par\vspace{0.8cm}\noindent
Reduction 1 is safe because if we do not include $v$ in the vertex cover, then we need to include all it's neighbors, otherwise there will exist an edge ${v,x}$, where $x$ is a neighbor of $v$ which is not included in the vertex cover, both of whose end-points will not be present in the vertex cover. If we include all the neighbors of $v$, then we will exceed the total budget $K$, on size of vertex cover. Therefore, we need to include $v$. That is why, if the current budget $K$, on vertex cover is 0, or if $c(v) = R$ and $K_R = 0$, we can return that we are dealing with No-instance as we need to include $v$.
\par\vspace{0.7cm}\noindent
\textbf{Reduction 2.} If there is a vertex $v$, with greater than $K_R$ neighbors with color, $R$, then add $v$ to the vertex cover and return the instance $(T,K_R,K-1)$ if $c(v) = B$, return the instance $(T,K_R-1,K-1)$ if $c(v) = R$. If $K-1 <0$ or $c(v) = R$ and $K_R-1 < 0$, return No-instance.
\par\vspace{0.8cm}\noindent
Reduction 2 is safe because if we do not include $v$ in the vertex cover, then we need to include all it's neighbors. But in that case, we will have to include more than $K_R$ vertices with $R$ color, which is not allowed as our budget on $R$ colored vertices is at most $K_R$. Therefore, we need to include $v$. Now, if the current budget $K$, on vertex cover is 0, or if $c(v) = R$ and $K_R = 0$, we can return that we are dealing with No-instance as we need to include $v$.
\par\vspace{0.8cm}\noindent
Apply reductions 1 and 2 until they are no longer applicable. In the end, suppose we get $(T,K_R,K)$ as the reduced instance. Note that in this instance every vertex $v$ has at most $K$ neighbors, of which at most $K_R$ will be colored $R$. Now, consider the following simple observation.
\par\vspace{0.5cm}\noindent
\textbf{Observation:} Either $v$ or all it's neighbors is in the vertex cover.
\par\vspace{1em}\noindent
The above observation is trivially true by the definition of vertex cover. Let $T[u,K_R,K,r_u]$ be \texttt{True} if there exists a vertex cover of sub-tree rooted at $u$ with exactly $r_u \leq K_R$ colored $R$ and at most $K$ total vertices, and \texttt{False} otherwise. The base case is $T[u,K_R,K,r_u] = \text{\texttt{True}}$ if $u$ is a leaf. Now, we consider the following two cases:
\par\vspace{0.5cm}\noindent
\textbf{Case 1.} $v$ is in the vertex cover.
\par\vspace{1em}\noindent
In this case, we add $v$ to the vertex cover and reduce $K$ by 1. If $c(v) = R$, we reduce $K_R$ by 1. Now, we look at all the neighbors of $v$. Let $N(v) = \{t_1,t_2,\dots,t_l\}$ be the set of neighbors of $v$. Note that from Reduction 1, we can say that $l\leq K$. Now, we need to branch on all these neighbors, but we also have to take care that the budget on red vertices is respected while building up the solution recursively. Therefore, while branching on each of the sub-trees, rooted at $t_1,t_2,\dots,t_l$. Let $r_1,r_2,\dots,r_l$ be the number of red vertices included in vertex cover from each of these sub-trees respectively. We want to satisfy the following constraint:
\begin{align*}
    \centering
    r_1 + r_2 + \dots + r_l = K_R
\end{align*}
We can show that the number of ways to satisfy this constraint is $\binom{l+K_R -1}{K_R-1}$. Let $S=\{S_1,S_2,\dots,S_m\}$ be the set of solutions for the above constraint. Then: 
\begin{align*}
    \centering
    T[v,K_R,K,r_v] = \bigvee_{s\in S}\bigwedge_{u \in N(v)} T[u,K_R - \mathbbm{1}_{c(u) = R},K-1,s(u)]
\end{align*}
where $\mathbbm{1}_{c(u) = R}$ is 1 when $c(u) = R$ and $s(u)$ is the number of $R$ vertices selected in vertex cover for sub-tree rooted at $u$ in solution $s$. Upon branching, we need to compute $T[u,K_R - \mathbbm{1}_{c(u) = R},K-1,s(u)]$ for all neighbors $t_i$ of $v$ by setting $r_i$ value of each neighbor, corresponding to a solution of the above constraint. The solution to the above constraint will cover all cases for this case, in which a valid solution can be obtained.
\par\vspace{0.5cm}\noindent
\textbf{Case 2.} $v$ is not in the vertex cover.
\par\vspace{1em}\noindent
In this case, we need to include all the neighbors of $v$ in the vertex cover. Then, we will branch on the grandchildren of $v$ in a similar fashion as we did in Case 1. We will solve the above constraint problem on each of the $v$'s children. Note that by Reduction 1, we know that $v$ can have at most $K^2$ grandchildren and if $t_1,t_2,\dots,t_l$ are the grandchildren of $v$, $r_1,r_2,\dots,r_l$ is the number of $R$ colored vertices included in vertex cover from each of the sub-trees rooted at these neighbors respectively, we need to compute $T[t_i,K_R,K,r_i]$ for all grandchildren $t_i$ of $v$ by setting $r_i$ value of each grandchild, corresponding to each solution of the constraint:
\begin{align*}
    \centering
    r_1 + r_2 + \dots + r_l = K_R
\end{align*}
Again, the number of ways to satisfy this constraint is $\binom{l+K_R -1}{K_R-1}$. Let $S=\{S_1,S_2,\dots,S_m\}$ be the set of solutions for the above constraint. Then: 
\begin{align*}
    \centering
    T[v,K_R,K,r_v] = \bigvee_{s\in S}\bigwedge_{u \in N(v)}\bigwedge_{t \in N(v)} T[t,K_R-|N_R(v)|,K-|N(u)|,s(t)]
\end{align*}
where $N_R(v)$ is the set of $R$ colored neighbors of $v$ and $s(t)$ is the number of $R$ vertices selected in vertex cover for sub-tree rooted at $t$ in solution $s$.
\par\vspace{0.5cm}\noindent
In the end, we return $\bigvee_{s}T[r,K_R,K,s(r)]$ as the final solution. If any one of the branch in our bounded search tree results in successful search for desired vertex cover, then the returned result will be \texttt{True}, and \texttt{False} otherwise.

\subsubsection{Time Complexity Analysis}
For each vertex in the tree, we branch into two possibilities. Corresponding to each vertex $v$, in the tree, we will compute $T[v,K_R,K,r_v]$ only once and memoize the result. Using dynamic programming will help us to not recompute the solution to same subproblems repeatedly. Now, the number of possibilities for $r_v$ is bounded by the number of solutions for our constraint, which is bounded by $\binom{l+K_R -1}{K_R-1}$ where $l$ is the number of grandchildren of $v$. We ignore the bound provided by Case 1 as that would be a lower order term in the analysis. Now, the number of grandchildren from our Reduction 1 is bounded by $K$. Therefore, corresponding to each vertex $v$, we need to at most compute $\binom{K^2+K_R -1}{K_R-1}$ values in the $T$ table.
\begin{align*}
    \centering
    \binom{K^2+K_R -1}{K_R-1} \leq (K^2+K_R)^{K_R}
\end{align*}
Computing each of these entries is an $\mathcal{O}(1)$ operation. Since we need to compute these values for every vertex, there will be $|V|$ such computations. Therefore, the overall time complexity of the presented algorithm is $\mathcal{O}(|V|(K_R+K^2)^{K_R})$.

\section{Miscellaneous}
An interesting problem which occurred to us when solving the above problems is \textproc{Vector Weights Vertex Cover}. It is defined as follows.\\
\newline
\fbox{
\parbox{13.5cm}{
	\textproc{Vector Weights Vertex Cover}\\
	\textsf{\bfseries Instance:} A graph, $G = (V,E,w,d)$, where function $w:V \rightarrow \mathbb{Q}^d$ and a parameter $k$, $k \in {Q}^d$.\\
	\textsf{\bfseries Compute:} A set $S \subseteq V$ such that $S$ is a vertex cover of $G$ and $\sum_{v\in S} w(v) \leq k$ where $\leq$ relation on vectors represents element-wise $\leq$ operation.\\
	\textproc{Bounded Number of Fractional Weights}\\
	\textsf{\bfseries Variation:} The problem is additionally parameterized with $l \in Z^d$, which contains the maximum number of vertices which have a fractional weight at the corresponding index.
}
}\\

\vspace{1em}
Note that the weights are vectors of real numbers, unlike in the $d = 1$ case, the weights can be negative here, since they cannot be trivially removed. It is due to the presence of the negative weights that the branching technique used above also fails. A few interesting questions can arise from the above problem. One such problem would be, ``For any $d$, is a reduction possible to a smaller $d$?'', ``If the answer to the above question is NO, is it useful to think of these as various hardness classes?''.

\bibliographystyle{plain}
\bibliography{input/report}

\begin{thebibliography}{1}

\bibitem{introtowvc}
Blow-ups, win/win’s and crown rules: some new directions in fpt, in:
  Proceedings of the 29th international workshop on graph theoretic concepts in
  computer science (wg’03), lec- ture notes in computer science, vol. 2880,
  2003, pp.1–12, kernelization algorithms for the vertex cover problem:
  theory and experiments, in: Proceedings of the workshop on al- gorithm
  engineering and experiments (alenex), new orleans, louisiana, january 2004,
  pp. 62–69.

\bibitem{inproceedings}
Faisal Abu-khzam, Rebecca L.~Collins, Michael Fellows, Michael Langston,
  W~Suters, and Christopher T.~Symons.
\newblock Kernelization algorithms for the vertex cover problem: Theory and
  experiments.
\newblock pages 62--69, 01 2004.

\bibitem{Carr:2000:RSC:338219.338271}
Robert~D. Carr, Srinivas Doddi, Goran Konjevod, and Madhav Marathe.
\newblock On the red-blue set cover problem.
\newblock In {\em Proceedings of the Eleventh Annual ACM-SIAM Symposium on
  Discrete Algorithms}, SODA '00, pages 345--353, Philadelphia, PA, USA, 2000.
  Society for Industrial and Applied Mathematics.

\bibitem{Chen2006ImprovedPU}
Jianer Chen, Iyad~A. Kanj, and Ge~Xia.
\newblock Improved parameterized upper bounds for vertex cover.
\newblock In {\em MFCS}, 2006.

\bibitem{Chlebik:2008:CRM:1330778.1331075}
Miroslav Chleb\'{\i}k and Janka Chleb\'{\i}kov\'{a}.
\newblock Crown reductions for the minimum weighted vertex cover problem.
\newblock {\em Discrete Appl. Math.}, 156(3):292--312, February 2008.

\bibitem{Cygan:2015:PA:2815661}
Marek Cygan, Fedor~V. Fomin, Lukasz Kowalik, Daniel Lokshtanov, Daniel Marx,
  Marcin Pilipczuk, Michal Pilipczuk, and Saket Saurabh.
\newblock {\em Parameterized Algorithms}.
\newblock Springer Publishing Company, Incorporated, 1st edition, 2015.

\bibitem{Kar72}
R.~Karp.
\newblock Reducibility among combinatorial problems.
\newblock In R.~Miller and J.~Thatcher, editors, {\em Complexity of Computer
  Computations}, pages 85--103. Plenum Press, 1972.

\bibitem{Raman2008}
Venkatesh Raman and Saket Saurabh.
\newblock Short cycles make w-hard problems hard: Fpt algorithms for w-hard
  problems in graphs with no short cycles.
\newblock {\em Algorithmica}, 52(2):203--225, Oct 2008.

\end{thebibliography}


\begin{thebibliography}{1}

\bibitem{Hartnell95}
B.L. Hartnell.
\newblock Firefighter! an application of domination.
\newblock {\em in: 25th Manitoba Conference on Combinatorial Mathematics and
  Computing}, 1995.

\end{thebibliography}
\end{document}